\documentclass[final]{arxiv}  

\usepackage{lineno,hyperref,graphicx,natbib}
\begin{document}

\doublespace

\addtolength{\voffset}{0.5in}

\title{\LARGE{Arecibo Radar Observations of Near-Earth Asteroid}}
\title{\LARGE{(3200) Phaethon During the 2017 Apparition}}
\author{Patrick A. Taylor$^{a, 1, *}$, Edgard G. Rivera-Valent\'{i}n$^{a, b}$, Lance A.M. Benner$^{c}$, Sean E. Marshall$^{a, 2}$, Anne K. Virkki$^{a, 2}$, Flaviane C.F. Venditti$^{a, 2}$, Luisa F. Zambrano-Marin$^{a, d, 2}$, Sriram S. Bhiravarasu$^{a, 1}$, Betzaida Aponte-Hernandez$^{a, 1}$, Carolina Rodriguez Sanchez-Vahamonde$^{a, 3}$, and Jon D. Giorgini$^{c}$}
\affil{}
\affil{$^{a}$Arecibo Observatory, Universities Space Research Association\\$^{b}$Lunar and Planetary Institute, Universities Space Research Association\\$^{c}$ Jet Propulsion Laboratory, California Institute of Technology\\$^{d}$Universidad de Granada}
\affil{}
\affil{Present Affiliations:  $^{1}$Lunar and Planetary Institute, Universities Space Research Association\\$^{2}$Arecibo Observatory, University of Central Florida, $^{3}$University of Western Ontario}
\affil{}
\email{ptaylor@usra.edu}

\journame{Planetary \& Space Science\\Special Issue on Asteroid (3200) Phaethon and Meteoroids}
\submitted{2018 November 15}
\revised{2018 December 16}
\accepted{2019 January 27}
\pubonline{2019 January 30}
\pubprint{}

\pages{16}
\tables{2}
\figures{5}

\quad\newline

\singlespace  

\clearpage

\noindent
NOTICE: This is the author's version of a work that was accepted for publication in Planetary \& Space Science. 
Changes resulting from the publishing process, such as peer review, editing, corrections, structural formatting, 
and other quality control mechanisms may not be reflected in this document. Changes may have been made to 
this work since it was submitted for publication.\\
\\
Publisher's copy:  \href{https://dx.doi.org/10.1016/j.pss.2019.01.009}{https://dx.doi.org/10.1016/j.pss.2019.01.009}\\

\noindent \textcopyright 2019. This manuscript version is made available under the CC-BY-NC-ND 4.0 license https://creativecommons.org/licenses/by-nc-nd/4.0/ 

\thispagestyle{empty}

\clearpage

\addtolength{\voffset}{-0.5in}


\setcounter{page}{1}

\noindent ABSTRACT:
We report Arecibo S-band (2380 MHz; 12.6 cm) radar observations of near-Earth asteroid (3200) Phaethon during the December 2017 apparition 
when Phaethon passed within 0.07 au of Earth.  Radar images with a resolution of 75 m per pixel reveal a roughly spheroidal shape more than 6 km 
in diameter at the equator with several discernible surface features hundreds of meters in extent.  These include a possible crater more than 1 km across 
located below 30$^{\circ}$ latitude and a roughly 600-m radar-dark region near one of the poles.  Overall, the radar images of Phaethon are reminiscent 
of those of (101955) Bennu, target of the OSIRIS-REx mission.  As such, the shape of Phaethon is suspected to have an equatorial ridge similar to 
the top-shaped models of several other radar-observed near-Earth asteroids as well as the optical images of (162173) Ryugu returned by the 
Hayabusa2 spacecraft.  Preliminary analysis of the radar data finds no satellites and gives no indication of a dusty coma at the time of these 
observations.


\section{Introduction}

Near-Earth asteroid (3200) Phaethon is well known for its very blue, B-type spectrum~\citep{gree85,binz01,leon10}, its brightening and bursts 
of dust activity near perihelion~\citep{jewi10,li13,jewi13}, and its dynamical link to the Geminid meteoroid stream~\citep{whip83,gust89,will93}.  
Because of its intriguing nature, Phaethon is a proposed flyby target of the Japanese Aerospace Exploration Agency (JAXA) Demonstration and 
Experiment of Space Technology for INterplanetary voYage Phaethon fLyby dUSt Science (DESTINY$^{+}$) mission~\citep{arai18} to better 
understand the origin and nature of interplanetary dust and the dust-ejection mechanism of active bodies.  The 0.069 au flyby on 2017 December 
16 was the closest Phaethon has come to Earth since its discovery in 1983~\citep{gree83}, which made the 2017 apparition the best opportunity 
to characterize this body with radar prior to the planned launch of DESTINY$^{+}$ in 2022.  Phaethon will not make a closer flyby of Earth until 2093. 

Using optical lightcurve data spanning over 20 years,~\citet{hanu16} produced a convex, three-dimensional shape model of Phaethon with a preferred 
spin pole at ($\lambda$, $\beta$) = (319$^{\circ}$, -39$^{\circ}$) $\pm$ 5$^{\circ}$ in ecliptic coordinates and a sidereal rotation period of 3.603958 
$\pm$ 0.000002 h.  The convex shape found by lightcurve inversion includes an equatorial ridge and has some similarity to the shapes of near-Earth 
asteroids observed by radar such as 66391 (1999 KW$_{4}$)~\citep{ostr06}, though the authors note the ridge on Phaethon in their shape model is 
more subdued and not as symmetric by comparison.  Thermophysical modeling of ground- and space-based infrared data by~\citet{hanu16} constrained 
the geometric visible albedo of Phaethon to 0.122 $\pm$ 0.008 and its effective diameter to 5.1 $\pm$ 0.2 km.  \citet{kim18} included additional optical 
data from the 2017 apparition and found a sidereal period in agreement with~\citet{hanu16}, a lightcurve amplitude of 0.075 $\pm$ 0.035 mag, a lower 
bound on the equatorial axis ratio of 1.07, and two preferred spin poles within 10$^{\circ}$ of ecliptic (308$^{\circ}$, -52$^{\circ}$) and (322$^{\circ}$, 
-40$^{\circ}$).  The convex shape found by lightcurve inversion has an equatorial axis ratio of 1.118 and, overall, is similar to that found by~\citet{hanu16} 
given the significant overlap in optical datasets used.  Both suggest the presence of an equatorial ridge and more uniformly sloped hemispheres than 
a sphere or oblate spheroid.  A non-convex, three-dimensional shape model by~\citet{kim18} suggests Phaethon has a significant concavity in one 
of its hemispheres, but the authors note a non-convex shape model made solely with optical lightcurves is not unique.

Here, we describe radar observations of Phaethon with the Arecibo planetary radar system during the 2017 apparition, which was the first planetary 
radar campaign after Hurricane Maria devastated the island of Puerto Rico.  These observations are a testament to the strength and tenacity of the 
people of Puerto Rico.  Radar images of Phaethon provide direct measurements of its size and surface features that have, to this point, only been 
inferred from optical lightcurve and infrared data.  Furthermore, the radar images are sufficient to produce a detailed non-convex, three-dimensional 
shape model, which is in progress and will be presented in a future manuscript.  Here, we concentrate on describing the general shape, spin state, 
reflectivity, near-surface structure, and topography of Phaethon and the search for a dusty coma and satellites.  This manuscript is organized as 
follows:  radar observations are described in Section 2, results in Section 3, and discussion and conclusions in Section 4.  

\section{Radar Observations}

A planetary radar observation of a near-Earth asteroid starts with the transmission of a monochromatic, circularly polarized signal for roughly the 
time it takes for light to reach and return from the target, the so-called round-trip time.  Before the echo returns to the transmitting station, the 
receiver is moved into focus and receives the echo in orthogonal circular polarizations for the same amount of time as the signal was transmitted.  
The cycling of transmitting and receiving is repeated for as long as the transmitting station can track the target.  

A typical signal transmitted from Arecibo is either unmodulated (continuous wave or cw for short) or modulated using a binary phase shift keying 
technique (ranging or imaging).  In the former case, the information received is only in terms of frequency due to Doppler shifting by the bulk motion 
and rotation of the target.  In the latter case, the phase of the transmitted signal is flipped or not flipped every baud, a unit of time of order one 
microsecond or less, where the pseudo-random series of phase flips has a code length after which the series repeats.  Correlation of the received 
signal with the transmitted code produces a delay resolution equal to the baud of the code, which translates to range resolution along the line of 
sight via multiplying the baud by the speed of light divided by two (due to the signal going out and returning).  One may sample the received signal 
faster than the baud value to make the range resolution finer, though the information contained in samples separated by less than one baud will 
be correlated.  The code length is chosen such that the bandwidth of the code, defined as the reciprocal of the product of the baud and the code 
length, is at least a factor of a few wider than the bandwidth of the echo from the target (see Section 3.1).  Using different code lengths for the same 
baud also allows for unambiguous range determination.  The resultant mapping of the received signal to time delay (or range) and frequency 
(Doppler shift) produces a delay-Doppler or range-Doppler radar image.  As the radar image is a projection of a three-dimensional object into a 
two-dimensional phase space, there is an inherent north-south ambiguity to radar images where more than one point on the object can map to the 
same range and Doppler position in the radar image.  This ambiguity can be broken with sufficient data from different aspect angles such that the 
same points on the object are not always ambiguous with each other.

The Arecibo planetary radar system observed Phaethon from 2017 December 15 to 19, including through its close approach of 0.069 au on 
December 16 and until it moved to negative declinations beyond the Arecibo field of view.  Radar observations from Arecibo are summarized 
in Table~\ref{tab:rad} and examined in detail in the following section.  With only one klystron amplifier in operation rather than the nominal two, the 
transmitted power of the planetary radar system was halved to roughly 400 kW.  Damage to the alignment of the reflecting surface by the winds of 
Hurricane Maria also reduced the telescope sensitivity at 2.38 GHz by at least 20\% overall and further as a function of azimuth and zenith angle.  
Nonetheless, this radar dataset provides the best information yet for physically characterizing the enigmatic Phaethon.

\begin{table*}[!h]
\begin{center}
\footnotesize
\begin{tabular}{ccccccccccccr}
\hline
UT Date & RA & Dec & Eph & $\Delta$ & P$_{\rm TX}$ & Baud & Spb & $\Delta$r &  $\Delta$f & Code & Start-Stop & Runs\\
 & ($^{\circ}$) & ($^{\circ}$) &  & (au) & (kW) & ($\mu$s) &  & (m) & (Hz) &  & hhmmss-hhmmss & \\
\hline
2017 Dec 15 & 22 & +36 & 609 & 0.071 & 363 & cw & - & - & 1.0 & none & 233607-234647 & 5\\
            &  &  &       &  & 365 & 4 & 2 & 300 & 1.9 & 1023 & 235233-005042 & 25\\
            &  &  &       &  & 375 & 4 & 2 & 300 & 1.9 & 8191 & 005248-005351 & 1\\
\hline
2017 Dec 16 & 8 & +28 & 609 & 0.069 & 379 & cw & - & - & 1.0 & none & 220223-221236 & 5\\
            &  &  &       &  & 393 & 1 & 2 & 75 & 1.0 & 8191 & 221909-002640 & 56\\
\hline
2017 Dec 17 & 356 & +19 & 611 & 0.071 & 370 & cw & - & - & 1.0 & none & 210239-211310 & 5\\
            &  &  &       &  & 408 & 1 & 2 & 75 & 1.0 & 8191 & 211903-234105 & 56\\
\hline
2017 Dec 18 & 346 & +9 & 613 & 0.077 & 393 & 1 & 2 & 75 & 1.0 & 8191 & 203407-212717 & 21\\
            &  &  &       &  & 374 & cw & - & - & 1.0 & none & 213130-214304 & 5\\
            &  &  &       &  & 394 & 1 & 2 & 75 & 1.0 & 8191 & 214726-224625 & 23\\
\hline
2017 Dec 19 & 339 & +2 & 613 & 0.087 & 370 & cw & - & - & 1.0 & none & 202701-203956 & 5\\
            &  &  &       &  & 391 & 2 & 4 & 75 & 1.0 & 1023 & 204616-214615 & 21\\
\hline
\end{tabular}
\end{center}
\caption{Radar observations of Phaethon with the Arecibo planetary radar system.  UT Date is the universal-time date on which the observation 
began. RA and Dec are the right ascension and declination of date, in degrees, of the lines of sight to Phaethon. Eph gives the Jet Propulsion Laboratory 
orbit solution number used from the Horizons ephemeris service.  $\Delta$ is Phaethon's distance from Earth in au; $\Delta$ multiplied by 1000 
approximately gives the round-trip time to the target. P$_{\rm TX}$ is the transmitted power in kilowatts.  Baud is the time (delay) resolution of the 
pseudo-random code transmitted; baud does not apply to unmodulated cw data.  Spb is the number of samples per baud recorded, which relates to the 
sampling rate as Spb/Baud.  $\Delta r$ is the range resolution of the resulting radar images.  $\Delta f$ is the frequency resolution of the processed data.  
Code is the length (in bauds) of the pseudo-random code transmitted.  The timespan of the received data is listed by the UT start and stop times.  Runs 
is the number of completed transmit-receive cycles.}
\label{tab:rad}
\end{table*}

\section{Results}

\subsection{Doppler-Only Spectra}

\begin{figure}[!h]
\begin{center}
\includegraphics[scale=0.45]{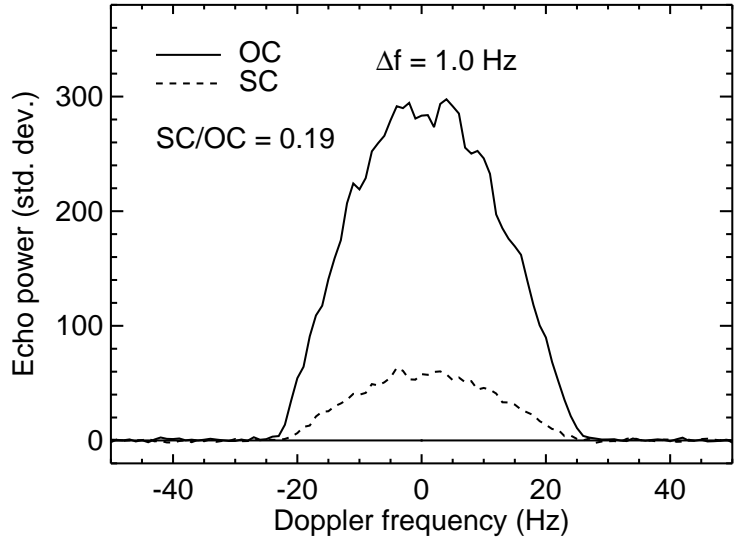} 
\end{center}
\caption[Doppler-only echo power spectra of (3200) Phaethon]{
Weighted sum of 25 Doppler-only echo power spectra of Phaethon from 2017 December 15 through 19 using the Arecibo planetary radar system.  
Echo power on the vertical axis is measured in standard deviations of the background (off-echo) noise.  The frequency resolution $\Delta$f on the 
horizontal axis is 1.0 Hz.  The bandwidth of the summed spectrum is $\sim$50 Hz, which is greater than the bandwidth on any given day due to 
small offsets in the predicted Doppler position of the echo between ephemeris solutions 609, 611, and 613.  Echo strengths in both polarizations 
are shown:  opposite circular (relative to transmission; OC) and same circular (SC).}
\label{fig:cw}
\end{figure}

As the target rotates, one limb approaches the observer while the other recedes such that the transmitted monochromatic signal is Doppler 
broadened according to the rotation rate of the target as projected along the observer's line of sight (see Fig.~\ref{fig:cw}).  Consequently, the 
radar echo of the target has a bandwidth $B$ given by:

\begin{equation}
B = \frac{4\pi\,D(\phi)}{\lambda P}\,\cos\,\delta
\label{eq:bw}
\end{equation}

\noindent where $D(\phi)$ is the diameter (breadth) of the target presented to the observer at rotation phase $\phi$, $\lambda$ is the wavelength 
of the transmitted signal, $P$ is the rotation period of the target, and $\delta$ is the sub-radar latitude on the target as viewed by the observer.  
For a sphere with a known diameter and rotation period, one can constrain the position of the spin axis from the variation in the observed bandwidth 
as the target moves across the sky.  The spin axis can only reside in certain regions on the celestial sphere such that the change in sub-radar latitude 
matches the change in echo bandwidth.  

Taking Phaethon to be roughly spherical, as implied by the low-amplitude optical lightcurves in the literature, the fact that the echo bandwidth stayed 
within the range of 45 to 49 Hz (Fig.~\ref{fig:cw}) over five days of observations from Arecibo suggests the sub-radar latitude, or at least its cosine, 
did not change considerably as Phaethon traversed more than 50$^\circ$ across the sky (see Table~\ref{tab:rad}).  The maximum bandwidth of 49 
Hz was observed on December 16 such that the sub-radar point was likely closest to the equator of Phaethon during its closest approach to Earth.  
We note that a sphere of diameter 5.1 km as found by~\citet{hanu16} rotating in 3.6 h would have a maximum possible bandwidth of only 39 Hz, 
meaning the equatorial dimensions of Phaethon must be larger than previously thought to match the observed bandwidths.  While the 3.6-h rotation 
period of Phaethon is well established, we also note that radar confirms such a rapid rotation is required to match the observed bandwidths; slower 
periods would result in an even larger size discrepancy from 5.1 km.

\begin{figure}[!h]
\begin{center}
\includegraphics[scale=0.45]{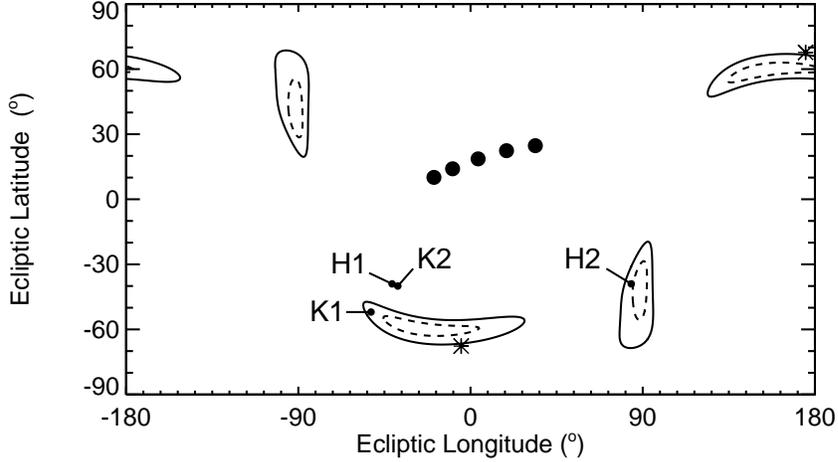} 
\end{center}
\caption[Pole search for (3200) Phaethon]{
Using a 6.2-km diameter sphere and a 3.6-h rotation period for Phaethon, the observed echo bandwidths from Arecibo radar observations in 2017 
are best fit by spin poles in four regions given in ecliptic coordinates.  Contours indicate chi-squared values 25\% (dashed) and 100\% (solid) greater 
than the global minimum reduced $\chi^{2}$ value of 0.76.  Large, filled circles indicate Phaethon's position on the sky during observations (see 
Table~\ref{tab:rad}) and asterisks indicate the heliocentric orbit normal (prograde) and anti-normal (retrograde).  H1, H2, K1, and K2 refer to the poles 
reported by~\citet{hanu16} and~\citet{kim18}.  Longitudes are translated from [0$^{\circ}$, 360$^{\circ}$] to [-180$^{\circ}$, 180$^{\circ}$] for clarity.}
\label{fig:pole}
\end{figure}

If one assumes a spherical shape for Phaethon and a rotation period of 3.6 h in Eq.~\ref{eq:bw}, but not a diameter, the observed bandwidths then 
place a joint constraint on the diameter and the projection due to the sub-radar latitude.  Comparing to the observed echo bandwidths for each possible 
spin pole over a range of assumed diameters, we find that the best-fit spherical diameter for Phaethon is about 6.2 km.  For the best-fit diameter, the 
spin pole likely falls in one of four regions shown in Fig.~\ref{fig:pole} such that Arecibo always viewed Phaethon within 20$^{\circ}$ of its equator; in 
other words, the radar lines of sight were always 90$^{\circ}$ $\pm$ 20$^{\circ}$ from the spin axis of Phaethon.  We caution that using the true shape 
of Phaethon rather than a sphere could cause the best-fit poles to shift by several degrees.

Recently published pole estimates by~\citet{hanu16} and~\citet{kim18} from optical lightcurve inversion are all retrograde with a cluster in the 
region of ecliptic longitude and latitude ($\lambda$, $\beta$) = (300$^{\circ}$ to 330$^{\circ}$, -35$^{\circ}$ to -55$^{\circ}$) and Hanu\v{s} et al.'s 
alternative pole at (84$^{\circ}$, -39$^{\circ}$).  One of the four regions that satisfy the Doppler-only data from Arecibo in 2017 is centered on 
(331$^{\circ}$, -59$^{\circ}$), about 20$^{\circ}$ from this cluster of poles and within 15$^{\circ}$ of parallel to the heliocentric orbit normal (165$^{\circ}$ 
obliquity) of Phaethon, which is (175$^{\circ}$, 68$^{\circ}$).  The other retrograde region is centered less than 5$^{\circ}$ from Hanu\v{s} et al.'s 
alternative pole; the other two regions are prograde mirrors of the retrograde solutions.  While the prograde solutions satisfy the Doppler-only data 
presented here due to the north-south ambiguity of radar, it is shown by~\citet{hanu16} and~\citet{kim18} that prograde solutions are incompatible 
with optical lightcurve data.

The received echo is recorded in both the opposite circular (OC) and same circular (SC) polarizations as transmitted, where the OC signal is expected 
for reflection from a plane mirror and the SC signal indicates multiple scattering from a rougher surface.  The SC/OC ratio is then a zeroth-order 
gauge of near-surface roughness at the wavelength scale of the radar, \textit{i.e.}, decimeter-scale roughness within a few meters of the surface.  In 
some cases, the SC/OC ratio is diagnostic of composition:  SC/OC values of order unity are often observed for E-type asteroids and SC/OC 
values $\sim$0.6 are often observed for V-type asteroids, while SC/OC values among the C and S complexes range from roughly 0.1 to 0.55~\citep{benn08} 
and are indistinguishable with radar alone.  The polarization ratio of Phaethon ranges from 0.18 to 0.20 $\pm$ 0.02 over the five days of observation 
with a weighted average of 0.19 $\pm$ 0.01 (Fig.~\ref{fig:cw}).  Though not elevated enough to be diagnostic of composition on its own, this polarization 
ratio is consistent with the B-type (C complex) composition determined from visible and near-infrared observations.  We note that~\citet{benn08} reported 
a value of 0.25 $\pm$ 0.02 from observations of Phaethon in 2007, though we caution that this value appears to come from a single spectrum compared 
to the 25 spectra summed in Fig.~\ref{fig:cw}.  It is not uncommon to see some variation in the SC/OC ratio of individual spectra.

The polarization ratio of 0.19 for Phaethon is very similar to the value of 0.18 $\pm$ 0.03 for (101955) Bennu~\citep{nola13}, also a B-type asteroid.  
Other B-type near-Earth asteroids observed with radar:  7753 (1988 XB)~\citep{binz04} and 153591 (2001 SN$_{263}$)~\citep{pern14,beck15} have 
published polarization ratios of 0.18 $\pm$ 0.02~\citep{benn08} and 0.17 $\pm$ 0.03~\citep{beck15}, respectively.  Though small-number statistics 
with only four examples thus far, there is a curious clustering of the indentified B-type near-Earth asteroids about a polarization ratio of 0.18.  

The polarization ratio of Phaethon is comparatively less than the polarization ratios of S-complex asteroids visited by spacecraft:   0.23 $\pm$ 0.03 for 
(4179) Toutatis (reported from archival data by~\citet{nola13}), 0.26 $\pm$ 0.04 for (25143) Itokawa~\citep{ostr04}, and 0.28 $\pm$ 0.06 for (433) 
Eros~\citep{magr01}.  Though not strictly statistically significant beyond the two-sigma level, the lower polarization ratio of Phaethon may indicate a 
somewhat smoother surface at decimeter scales than observed by the Hayabusa and NEAR Shoemaker spacecraft as they landed on Itokawa and 
Eros, respectively.  While Chang'e-2 found the surface of Toutatis to be globally smoother than Itokawa~\citep{huan13} to the eye, the flyby images of 
Toutatis had resolution coarser than $\sim$3 m per pixel; this is more than an order of magnitude larger than the 12.6-cm wavelength of the Arecibo 
planetary radar system such that the roughness scales probed by the flyby images and the radar observations are not comparable.

The total radar cross section of Phaethon (OC plus SC) is 2.1 km$^{2}$ with a calibration uncertainty of order 35\% (allowing for additional uncertainty 
due to the damage from Hurricane Maria), meaning a perfect, isotropically reflecting metal sphere with a 2.1-km$^{2}$ geometric cross section (1.6-km 
diameter) would return the same signal strength as observed from Phaethon under the same observing circumstances.  If Phaethon were a sphere of 
diameter 6.2 km, its OC radar albedo would be about 6\%, placing it among the lowest values compiled for near-Earth asteroids~\citep{nees12} and 
more akin to comets, which range from about 4\% to 10\%~\citep{harm99,harm04,harm11}.  For comparison, the radar albedo of Bennu is 12\% and 
Itokawa, Toutatis, and Eros range from 14\% to 25\%.  Such a low radar albedo indicates high porosity or low bulk density below 1 g/cm$^{3}$ near 
the surface~\citep{magr01,harm04}.

The 12.6-cm wavelength of the Arecibo planetary radar system is also sensitive to a dusty coma of particles larger than a few cm in diameter.  
Detection of comae when observing comets with radar is common, two of the strongest examples being C/IRAS-Araki-Alcock~\citep{harm89} 
and 130P/Hartley 2~\citep{harm11}.  These comets show a ``peak and wings" pattern in Doppler-only echo power spectra characterized by a 
bright, narrow echo from the solid nucleus surrounded by a skirt up to hundreds of Hz wide caused by the outflow of particles in the dusty coma.  
For Phaethon, there is no evidence of such a skirt in Fig.~\ref{fig:cw} as the noise background is at a constant level in a 2 kHz bandwidth about 
the echo.  Concurrent optical observations on 2017 December 17 at an effective wavelength of 585 nm with the Hubble Space Telescope by~\citet{jewi18} 
similarly show no evidence for a dust tail with optical depth greater than 3\,$\times$\,10$^{-9}$.

Satellites of near-Earth asteroids are sometimes detected in Doppler-only echo power spectra as a strong, narrow spike superimposed upon the 
broader echo of the primary, \textit{e.g.}, 185851 (2000 DP$_{107}$)~\citep{marg02,naid15} and 2001 SN$_{263}$~\citep{beck15}.  The narrow 
bandwidth of the satellite is a combination of its smaller size and longer rotation period, typically tidally locked to its orbital period, relative to the 
larger, more rapidly rotating primary.  However, we find no such evidence for satellites as no conspicuous spikes are noted in the echo power 
spectra of Phaethon that comprise Fig.~\ref{fig:cw}.

\subsection{Range-Doppler Images}

\begin{figure*}[!h]
\begin{center}
\includegraphics[scale=0.24]{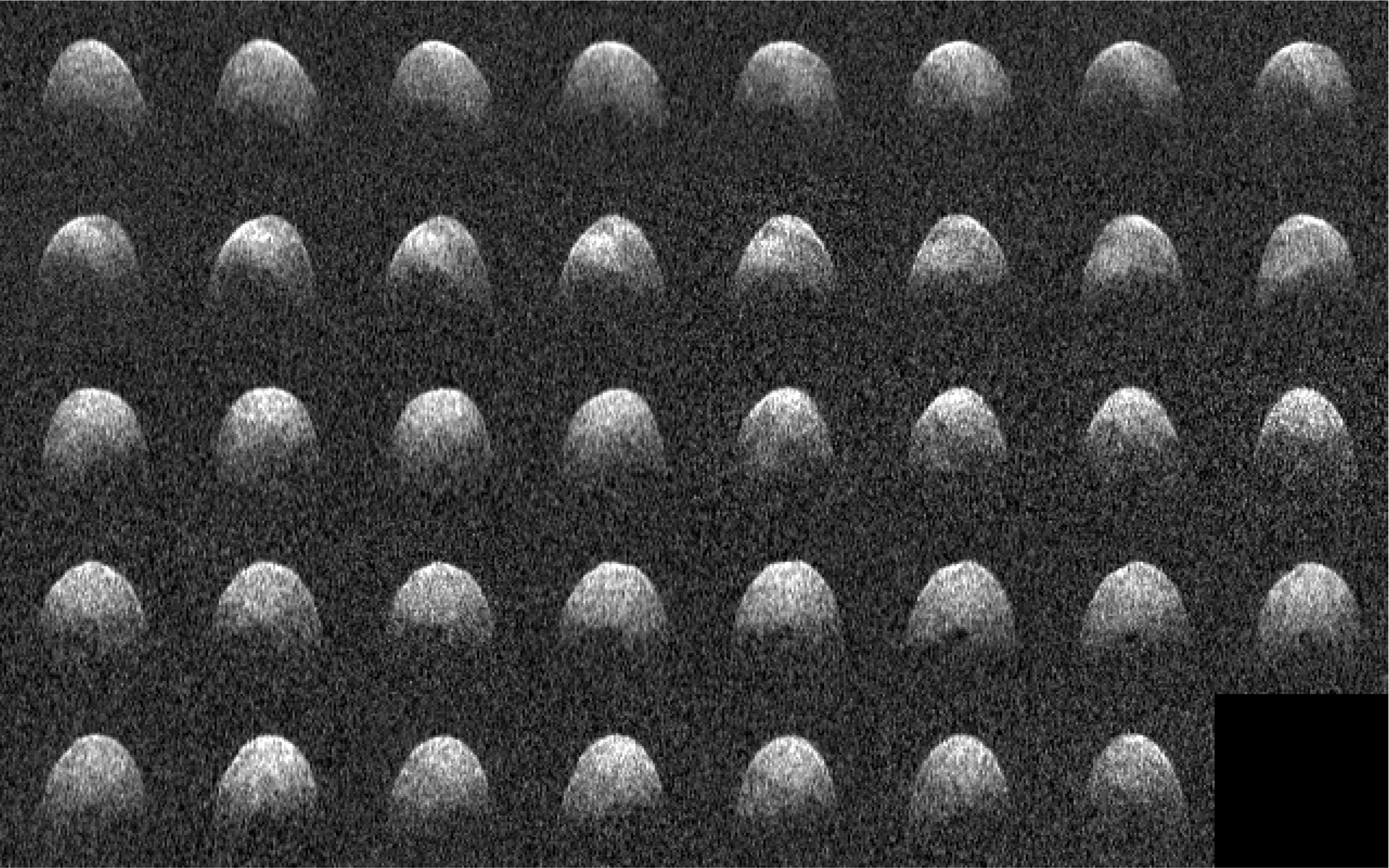}
\end{center}
\caption[Radar image mosaic of Phaethon]{
A subset of range-Doppler images of Phaethon from 2017 December 16, 17, and 18 showing a complete rotation.  Each frame is the sum of two 
consecutive transmit/receive cycles and has range (distance from the observer) increasing downward at 75 m per pixel and Doppler frequency 
increasing to the right at 1.0 Hz per pixel.  Frames are read left to right then top to bottom such that Phaethon appears to rotate counterclockwise
with each frame having $\sim$6$^{\circ}$ of rotational smear.  The asymmetry in the leading edge of the echo is apparent in the first column.  Two 
distinct surface features are visible in several frames:  a candidate concavity behind the leading edge in the second row of images and a radar-dark 
spot near the trailing edge in fourth row of images.  Contrast and brightness are enhanced to increase the visibility of features.}   
\label{fig:images}
\end{figure*}

Radar images (Fig.~\ref{fig:images}) confirm that Phaethon is spherical to zeroth order, but show evidence for equatorial elongation, an equatorial
ridge, and uniformly sloped hemispheres as found from lightcurve inversion~\citep{hanu16,kim18} and found for several other near-Earth asteroids 
observed with radar.  The echo depth measured from the leading to trailing edge in the radar images is at least 3 km.  If we are seeing the entire leading 
hemisphere, the equatorial diameter is then at least 6 km.  Similarly, by Eq.~\ref{eq:bw}, the maximum observed bandwidth of 49 Hz corresponds 
to a maximum breadth of about 6.4 km.  The leading edge of Phaethon deviates from the semi-circular edge expected for a sphere, appearing 
somewhat asymmetric in the radar images, \textit{e.g.}, column 1 of Fig.~\ref{fig:images}, and tapering from broad to narrow, \textit{e.g.}, row 3 of 
Fig.~\ref{fig:images}, and back again.  The bandwidths of the radar images vary with rotation at about the 5\% level corresponding to slightly oblong 
equatorial dimensions of 6.4 by 6.1 km.  Such an elongation of the equatorial region is consistent with a low-amplitude lightcurve, though it is below 
the lower limit of 7\% determined by~\citet{kim18} and significantly less elongated than the 12\% value of their convex shape model found by 
lightcurve inversion.  A mean equatorial diameter of 6.25 km is also consistent with the simple spherical model used to fit the Doppler-only echo 
power spectra in the previous section.  We note that such a diameter is significantly larger than the 5.1 km effective diameter found via thermophysical 
modeling by~\citet{hanu16}, though the explanation for this discrepancy of 20\% goes beyond the scope of this preliminary study.

\begin{figure*}[!h]
\begin{center}
\includegraphics[scale=0.45]{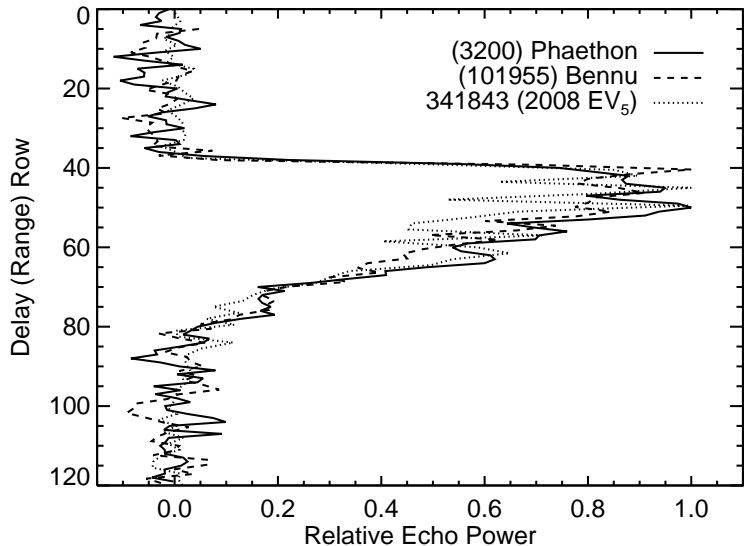}
\end{center}
\caption[Evidence for a ridge on Phaethon]{
The distribution of echo power as function of echo depth, made by collapsing the horizontal dimension of the range-Doppler image, from the first 
frame of Fig.~\ref{fig:images} suggests Phaethon has a shape similar to Bennu or 2008 EV$_{5}$, which are known to have equatorial ridges.  The 
images of Bennu and 2008 EV$_{5}$ were scaled to have the same echo depth as Phaethon for ease of comparison.  Signal in at least 40 rows at 
75 m per pixel suggests Phaethon is at least 3 km in visible extent or at least 6 km in diameter at the equator.} 
\label{fig:ridge}
\end{figure*}

The consistency of the echo power, shown by the gradual, rather than sharp, decrease in the brightness of pixels from the leading to trailing 
edge of the echoes in Fig.~\ref{fig:images}, especially the very first frame, also suggests that Phaethon is not quite spherical.  For a sphere, 
the echo power decreases sharply with echo depth as the incidence angle increases with latitude from the equator to the pole.  On the other 
hand, a shape with a roughly constant angle of incidence as a function of latitude due to uniformly sloped hemispheres will have stronger echo 
power with increasing latitude than a sphere.  \citet{busc11} compared the echo-power distribution expected from a sphere to the echo-power
distribution observed from near-Earth asteroid 341843 (2008 EV$_{5}$) to show that a shape with an equatorial ridge and roughly uniformly
sloped hemispheres, sometimes referred to as a top, diamond, biconal, or bipyramidal shape, provides a much better fit to the radar data than a 
simple sphere.  Fig.~\ref{fig:ridge} shows the relative echo power as a function of echo depth, in other words the sum of the echo power in each 
row of the radar image scaled to the brightest row, for representative range-Doppler images of Phaethon, Bennu, and 2008 EV$_{5}$.  The results 
are strikingly similar with a sharp increase in echo power at the leading edge of the echo and a gentle, almost linear, decrease in echo power with 
echo depth, suggesting Phaethon has a shape akin to Bennu~\citep{nola13} or 2008 EV$_{5}$.

At first glance, Phaethon is rather nondescript in radar images with a resolution of 75 m per pixel, suggesting the surface of Phaethon is relatively 
smooth on the scale of hundreds of meters.  With the exception of two major features, described below, there is not much contrast in radar brightness
between different regions on Phaethon.  In the first row of images, the left side of each echo tends to be slightly darker than the right side; the second 
row shows the most contrast due to the first major feature, while the remaining images have only subtle contrast across the echo.  

\begin{figure*}[!h]
\begin{center}
\includegraphics[scale=0.49]{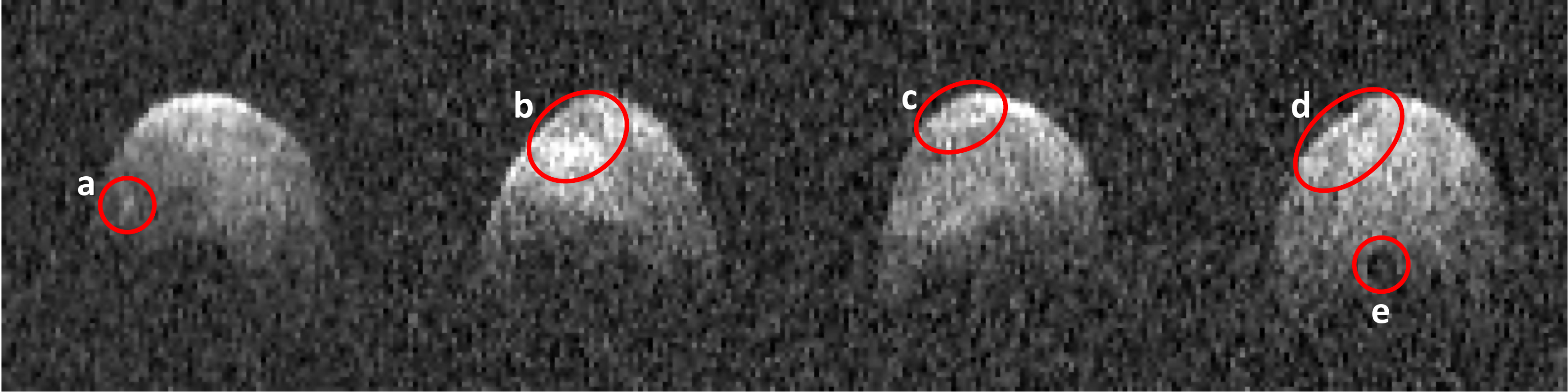}
\end{center}
\caption[Radar image mosaic of Phaethon]{
Enlargements of individual range-Doppler images from Fig.~\ref{fig:images} showing the main surface features of Phaethon described in the text and 
summarized in Table~\ref{tab:feat}:  (a) candidate 300-m boulder or raised region, (b) candidate concavity at low latitude, (c) candidate concavity/ridges 
near the equator, (d) linear, km-scale facet or ridge near the equator, and (e) 600-m radar-dark spot near one of the poles.  Contrast and brightness are 
enhanced and speckle noise is reduced to increase the visibility of the features.} 
\label{fig:enhanced}
\end{figure*}

The first major feature, found in the second row of Fig.~\ref{fig:images}, is a radar-bright region adjacent to a darker region that rotates counter-clockwise 
from the center of the frame to the lefthand limb.  This feature, enlarged in Fig.~\ref{fig:enhanced} and labeled as (b), is more than 1 km in diameter.  
The bright leading edge transitions to a radar-dark region followed by a radar-bright region that finally transitions to the more global brightness of the 
echo.  While this could indicate a region of radar-albedo variation, a more likely explanation is a topographical low such as a concavity or crater below 
30$^{\circ}$ latitude.  Compared to otherwise flat topography, roughly half the concavity would have a more extreme angle of incidence, while the other 
half would have a less extreme angle of incidence than the surrounding area, which could result in the observed variation in radar brightness.  

\begin{table*}[!b]
\begin{center}
\begin{tabular}{clccc}
\hline
& Surface Feature & Sub-Observer Time & Size & Latitude\\
\hline
a & Candidate Boulder & 2017 Dec 17, 22:13 UT & 0.3 km & $\sim$20$^{\circ}$ \\  
b & Candidate Concavity & 2017 Dec 18, 20:59 UT & $>$1 km & $<$ 30$^{\circ}$ \\
c & Candidate Concavity & 2017 Dec 18, 21:27 UT & 0.5-1.5 km & $<$ 10$^{\circ}$ \\
d & Linear Facet & 2017 Dec 17, 21:30 UT & 2 km & $\sim$0$^{\circ}$ \\
e & Polar Dark Spot & 2017 Dec 17, 21:35 UT & 0.6 km& $>$ 80$^{\circ}$ \\
\hline
\end{tabular}
\end{center}
\caption{Surface features of Phaethon as described in the text.  Sub-observer time is when the feature was closest to zero Doppler shift (horizontally 
centered) in the range-Doppler images of Fig.~\ref{fig:images}.  Features observed on different dates are consequences of rotation phase and 
illumination geometry; not all features were evident on all dates.  Note that extracting the true longitudes and latitudes of the features and extrapolating 
the sub-observer times to other epochs requires knowledge of the true shape and spin state of Phaethon to properly account for apparent rotation as 
Phaethon moved across the sky.}
\label{tab:feat}
\end{table*}

The second conspicuous feature is a dark spot (an absence of radar reflection) roughly 600 m in diameter near one of Phaethon's poles, seen at 
the trailing edge of the radar images in the fourth row of Fig.~\ref{fig:images} and enlarged in Fig.~\ref{fig:enhanced} as feature (e).  This dark spot 
is likely a surface patch that is flatter than its surroundings; in other words, the incidence angle of the radar beam is essentially grazing the surface 
causing minimal reflection from the patch.  Similarly, it could be a concavity or a radar shadow cast by a ridge near one of the poles that prevents the 
radar signal from illuminating the pole itself.  Faint echoes from near the poles have been observed for other near-Earth asteroids such as the primary 
component of triple system 136617 (1994 CC)~\citep{broz11} and for 2008 EV$_{5}$~\citep{busc11}, though not with the same striking contrast as 
found on Phaethon.  Another, but less likely, explanation is a region of significantly lower radar albedo near one of Phaethon's poles compared to the 
rest of the surface.  While only clearly visible in a few frames of Fig.~\ref{fig:images}, this is the only feature consistently found in the sums of all 
images from each day.  The position of this feature in the radar images also moves closer to the observer each day suggesting the sub-radar point 
on Phaethon was moving to higher latitudes, which is supported by the spin-axis constraints in the previous section.

Two less obvious features are found in the last column of Fig.~\ref{fig:images}.  In the last frame of the second row, left of center on the leading edge,
enlarged in Fig.~\ref{fig:enhanced} and labeled as (c), is a feature with similar brightness variation as the candidate concavity (b) described above.  
At the 75-m resolution of the images and the oblique projection along the leading edge, it is difficult to tell if this is a candidate concavity very near 
to the equator of Phaethon or ridge lines below about 10$^{\circ}$ latitude that cause the bright-dark-bright variation in the radar images.  The echo 
depth corresponds to at least 500 m between the leading edge and the bright ridge behind it, but the bandwidth is nearly a quarter of the entire echo, 
suggesting a breadth of roughly 1.5 km.

The other feature, visible throughout the fourth row of Fig.~\ref{fig:images} and enlarged in Fig.~\ref{fig:enhanced} and labeled as (d), is a linear facet 
along the leading edge up to 2 km in length based on its bandwidth.  In the fifth image in the fourth row, the facet is at the top of the image at the center 
of the leading edge, spanning nearly one third of the entire echo.  Most notably, in the last image in the fourth row, the feature is radar bright on the 
leading edge, but is radar dark immediately behind the leading edge, suggesting this facet along the equator is raised somewhat above its surroundings, 
possibly due to sharp topography along the equatorial ridge.  A similar pattern is subtle, but also noticeable elsewhere along the equator, including along 
the righthand limbs of the last images in the second and third rows in Fig.~\ref{fig:images} and to the right of feature (c) in Fig.~\ref{fig:enhanced}.

The final feature is a possible boulder about 300 m in extent in the lower-left region of the fifth frame of the first row of Fig.~\ref{fig:images} and also
enlarged in Fig.~\ref{fig:enhanced} and labeled (a).  This is the only frame where this feature is clearly visible suggesting that the grazing incidence 
angle of the radar beam near the trailing edge of the echo happened to illuminate the feature such that it produced a significantly brighter, more 
mirror-like radar return than its surroundings.  Other than this cluster of bright pixels, there is little evidence of large, hundred-meter scale candidate 
boulders or raised topography away from the equator of Phaethon.

Satellites of near-Earth asteroids are often discovered with radar because the components are unambiguously separated by the fine spatial 
resolution of radar images (\textit{e.g.}, \citet{marg02,ostr06,broz11,beck15}), even if not clearly visible in Doppler-only echo power spectra.  
Bright satellites are identifiable in individual radar images, while some fainter satellites are found in the sum of all images from a certain day by
the arcuate streak made as the satellite moves in its orbit about the primary component.  However, in this preliminary analysis, no obvious 
candidate satellites of Phaethon are noted on the scale of one hundred meters in diameter, neither in individual images nor the daily sums.  
Optical observations with the Hubble Space Telescope by~\citet{jewi18} similarly show no evidence for companions co-moving with Phaethon 
down to about 25 m in diameter assuming a Phaethon-like albedo of 0.12.

\section{Discussion and Conclusions}

Further analysis and thorough three-dimensional shape modeling of Phaethon are underway and are necessary to better inform the DESTINY$^{+}$ 
mission and to better understand Phaethon-like bodies.  Even without a complete shape model at this stage, we can deduce from the characteristics 
of the radar images that Phaethon's shape is similar to that of OSIRIS-REx target Bennu~\citep{nola13}, 2008 EV$_{5}$~\citep{busc11}, and
several primary components of binary near-Earth asteroids~\citep{ostr06,broz11,beck15,naid15}, whose shape models in turn are very similar to 
the optical images of (162173) Ryugu returned by the Hayabusa2 spacecraft~\citep{sugi18}.  Though Phaethon is more than an order of magnitude 
larger in diameter and three orders of magnitude larger in volume than Bennu, the smallest of those listed above, the general top shape with an 
equatorial ridge and uniformly sloped hemispheres appears common among near-Earth asteroids, independent of taxonomic type and composition.  

Taking Bennu as an example, if Phaethon is top-shaped rather than a 6.2-km sphere, we can expect its volume to be as much as 30\% less than 
the spherical assumption.  Its equivalent spherical diameter could then be as small as 5.5 km or less than 10\% larger than the value found by 
thermophysical modeling by~\citet{hanu16}.  While more manageable than a 20\% discrepancy, the reason for the difference in size determined 
from the radar and the optical and infrared datasets will need to be addressed by future work.  The extent of the volume difference compared to 
a sphere will also depend on how squashed the true shape is along the spin axis.  This dimension is typically the hardest to constrain due to the 
projection of the true three-dimensional shape to the two-dimensional range-Doppler radar images.  Constraints on this dimension during the 
shape-modeling process are often improved through the use of optical lightcurves, of which there are many for Phaethon, that depend on illuminated 
surface area.  An equivalent diameter of 5.5 km or more also requires revision of the geometric visible albedo from 0.122~\citep{hanu16} to 0.105 
or less.  Also, for a non-spherical shape, the cross-sectional area is expected to decrease resulting in a slightly higher OC radar albedo of $\sim$7\% 
and a near-surface bulk density of $\sim$1 g/cm$^{3}$.

The general shape and spin state of Phaethon reported here from the Arecibo radar data from 2017 are largely consistent with the low amplitude, 
optical lightcurves reported in the literature.  In addition, radar images confirm the rapid rotation period of $\sim$3.6 h and slight asymmetry in the 
equatorial region and support a spin axis $\sim$20$^{\circ}$ from the~\citet{hanu16} preferred pole and those of~\citet{kim18}.  Further refinement 
of the shape and spin state is underway using a combination of radar data from Arecibo and Goldstone in 2017, Arecibo in 2007 (albeit limited), 
and optical lightcurve data.  We expect only one of the four regions shown in Fig.~\ref{fig:pole}, the region near the clustering of Hanu\v{s} et al. 
and Kim et al.'s preferred poles, will satisfy the constraints of all available radar and optical data.  Though the Hanu\v{s} et al. alternative pole is 
consistent with our Doppler-only echo power spectra, the authors note that the alternative pole requires an unphysical thermal inertia to fit their 
infrared dataset.  The exact position of the spin axis will depend on the final non-convex, three-dimensional shape model of Phaethon.

Despite its rapid rotation and occasional dusty outbursts, we find no clear evidence in the radar data of Phaethon for satellites or a coma of 
cm-scale or larger particles.  Further analysis will be required to place upper limits on the size of possible satellites and the density of cm-scale 
particles around Phaethon.  We note that these radar observations took place when Phaethon was 1 au from the Sun and prior to its perihelion, 
which occurred on 2018 January 25, so it is not surprising that we find no clear evidence for activation.

\section*{Acknowledgements}

This research could not be done without the dedicated staff of the Arecibo Observatory, whose tireless work in the wake of Hurricane Maria's
devastation of Puerto Rico helped the Arecibo community recover and allowed these observations to be completed.  The authors thank the 
two anonymous reviewers for their helpful comments that clarified the manuscript.  Arecibo Observatory is a facility of the National Science 
Foundation operated under cooperative agreement, at the time of these observations, by SRI International in alliance with Universities Space 
Research Association (USRA) and Ana G. M{\'e}ndez Universidad Metropolitana (UMET) and is currently operated under cooperative agreement
No. 1822073 by the University of Central Florida (UCF) in alliance with Yang Enterprises, Inc. and UMET.  These observations with the Arecibo 
planetary radar system and related data analysis were supported by National Aeronautics and Space Administration (NASA) Near-Earth Object 
Observations (NEOO) Program grant NNX13AQ46G to USRA.  Work at Arecibo Observatory since 2018 April 1 was supported by NASA NEOO 
Program grant 80NSSC18K1098 to UCF.  Part of this work was performed at the Jet Propulsion Laboratory (JPL), California Institute of Technology, 
under contract with NASA.  This work made use of the JPL Horizons ephemeris service.  This work made use of NASA's Astrophysics Data System.

\bibliographystyle{model2-names.bst}
\bibliography{Phaethon-arXiv}

\end{document}